# Modeling the Crust and Upper Mantle in Northern Beata Ridge (CARIBE NORTE Project)

Diana Núñez1, Diego Córdoba1, Mario O. Cotilla1, Antonio Pazos2

[1] Dpto. Física de la Tierra, Astronomía y Astrofísica I (Geofísica y Meteorología), Universidad Complutense de Madrid, Avda. Complutense s/n, 28040 Madrid, Spain. E-mail: dianane@fis.ucm.es

[2] Real Observatorio e Instituto de la Armada, C/Cecilio Pujazón, s/n, 11100 San Fernando, Spain.

*Abstract*—The complex tectonic region of NE Caribbean, where Hispaniola and Puerto Rico are located, is bordered by subduction zone with oblique convergence in the north and by incipient subduction zone associated to Muertos Trough in the south. Central Caribbean basin is characterized by the presence of a prominent topographic structure known as Beata Ridge, whose oceanic crustal thickness is unusual. The northern part of Beata Ridge is colliding with the central part of Hispaniola along a transverse NE alignment, which constitutes a morphostructural limit, thus producing the interruption of the Cibao Valley and the divergence of the rivers and basins in opposite directions. The direction of this alignment coincides with the discontinuity that could explain the extreme difference between west and east seismicity of the island. Different studies have provided information about Beata Ridge, mainly about the shallow structure from MCS data. In this work, CARIBE NORTE (2009) wide-angle seismic data are analyzed along a WNW-ESE trending line in the northern flank of Beata Ridge, providing a complete tectonic view about shallow, middle and deep structures. The results show clear tectonic differences between west and east separated by Beata Island. In the Haiti Basin area, sedimentary cover is strongly influenced by the bathymetry and its thickness decreases toward to the island. In this area, the Upper Mantle reaches 20 km deep increasing up to 24 km below the island where the sedimentary cover disappears. To the east, the three seamounts of Beata Ridge provoke the appearance of a structure completely different where sedimentary cover reaches thicknesses of 4 km between seamounts and Moho rises up to 13 km deep. This study has allowed to determine the Moho topography and to characterize seismically the first upper mantle layers along the northern Beata Ridge, which had not been possible with previous MCS data.

## 1. Introduction

It is well known that the Caribbean–North America plate boundary zone is a complex region where different tectonic processes occur (Fig. 1a). This plate boundary zone (PBZ) consists of a 100–250-km-wide seismogenic area of mainly left-lateral strike-slip deformation, which extends over 2000 km along the northern edge of the Caribbean plate (CP) (MANN *et al.* 1995). The relative movement is towards North America plate in the E–NE direction, with a rate of 18–20 mm/yr and an azimuth of 70°, according to GPS measurements (MANN *et al.* 2002). These results explain the geodynamics of the Caribbean–North America plate boundary and imply maximum oblique convergence between both plates that are centered on Hispaniola Island. The area of such islands as Hispaniola and Puerto Rico is limited by a subduction zone with oblique convergence in the North and an incipient subduction zone associated to Muertos Trough in the South. The left-lateral Septentrional strike-slip fault (SFZ) and the south-dipping thrust faults of the North Hispaniola deformed belt accommodate deformation in the North, while in the South, the deformation is accommodated along the left-lateral Enriquillo–Plantain Garden Fault (EPGFZ) and the north-dipping thrust faults of Muertos deformed belt which is separated by the Beata Ridge. The Beata Ridge is a prominent NE–SW trending topographic structure in the central Caribbean basin, characterized by unusually thick oceanic crust (up to 20 km). It is believed to form a part of the ancient Caribbean oceanic plateau (RÈVILLON *et al.* 2000). It extends 400 km south of Cape Beata, Hispaniola, dividing the Caribbean into the Colombia and Venezuela basins (Fig. 1a), which are subducting normally below the South American deformed belt.

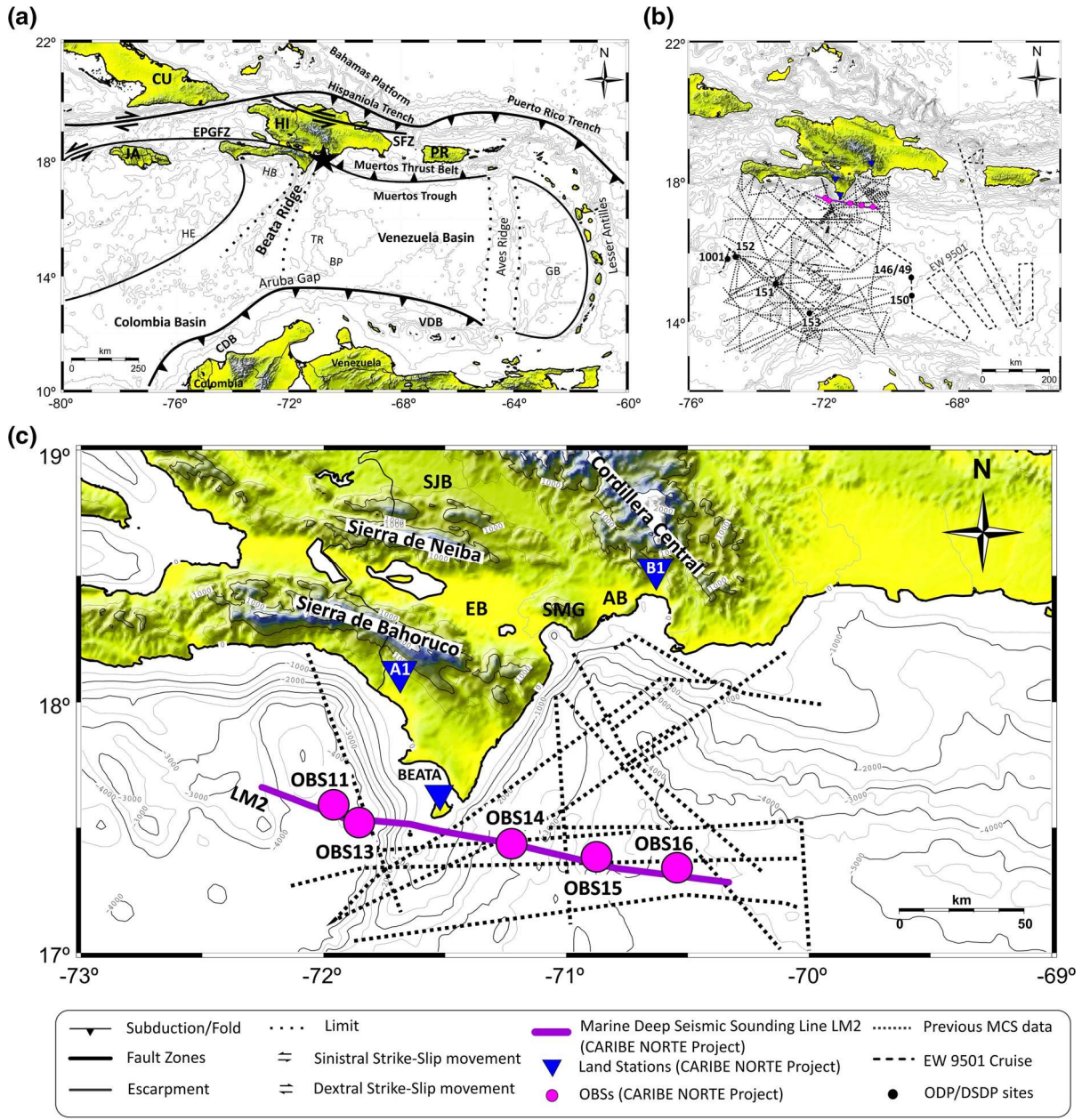

Figure 1

Deployment map with stations and seismic sources used in the CARIBE NORTE project and previous studies carried out in northern Beata Ridge region. a Tectonic frame of Caribbean region. *Black star* represents the position of the earthquake occurred on 1673, Ms = 7.5 (from COTILLA *et al.* 2007). b Previous studies in the area (MCS, SCS, DSDP and ODP data). *Dashed lines* represent the different track lines obtained from EW-9501 cruise (1995), MCS Casis seismic survey (1992), SeaCarib 1 cruise (1985) and other data provided by the Marine Geophysical Data Center. Black dots denote DSDP and ODP sites (1001, 146/49, 150, 151, 152, 153) (*Map* modified from MAUFFRET and LEROY 1999; and DIEBOLD *et al.* 1999). c Seismic wide-angle LM2 profile carried out during the CARIBE NORTE project. The Greater Antilles islands are represented by CU (Cuba), JA (Jamaica), HI (Hispaniola) and PR (Puerto Rico). *HE* Hess Escarpment, *HB* Haiti Basin, *TR* Taino Ridge, *BP* Beata Plateau, *GB* Grenada Basin, *CDB* Colombia Deformed Belt, *VDB* Venezuela Deformed Belt, *EPGFZ* Enriquillo–Plantain Garden Fault Zone, *SFZ* Septentrional Fault Zone, *SJB* San Juan Basin, *EB* Enriquillo Basin, *AB* Azu´a Basin, *SMG* Sierra de Mart´ın Garc´ıa

Despite the large number of studies conducted in the northern Caribbean plate [(EWING *et al.* 1960), (FOX *et al.* 1970), (TALWANI *et al.* 1977), (MOORE and FAHLQUIST 1976), among others], there are certain areas where there is no consensus on the geodynamic models that govern the structure and tectonic development of the zone due to the absence of in-deep geophysical research. In particular, the Beata Ridge has been ana- lyzed from North to South in different studies through Multi-Channel Seismic (MCS) and Single-Channel Seismic data, DSDP and ODP sites (MAUFFRET and LEROY 1999; DIEBOLD *et al.* 1999; DRISCOLL and DIE-

BOLD 1999). These studies produced shallow structure along several transversal lines obtained from different cruises (EW-9501, MCS Casis seismic survey, Seacarib 1, among others) (Figs. 1b, c).

## 2. Regional Setting

The Beata Ridge is a NE–SW trending structure located in the interior of the CP, between the extensive Colombian and Venezuelan basins. The Aruba Gap, a narrow connection between these two basins, truncates the Beata Ridge structure before it reaches the continental slope of South America. In contrast, the northern part of the Beata Ridge collides with the central part of Hispaniola in relation to the eastwards drift of the CP relative to Hispaniola. This collision results from the buoyancy of thick volcanic crust beneath Hispaniola that resists subduction (BURKE et al. 1982; BURKE 1988; MERCIER DE LEPINAY et al. 1988). The Beata Ridge and the central Caribbean were dredged (FOX et al. 1970) and drilled during the Deep Sea Drilling Program (DSDP) (Fig. 1b). The samples consisted of basalts (DSDP sites 150 and 151) and, or at certain sites, sediments with dolerites (DSDP site 152) (DONNELLY et al. 1973, 1990; DIEBOLD et al. 1999).

The northern part of the ridge is *100 km wide and emerged (Sierra de Bahoruco in the southern part of Hispaniola Island) and the southern part is approximately 3500 km wide and more than 4000 m below sea level, thus suggesting that northern part of the ridge is colliding with the E–W trending Caribbean island arc. This structure brings the Caribbean crust up to shallower depths, and it appears to be underlain by somewhat thicker crust than that of the adjacent Venezuelan Basin to the east. It also seems to be bounded on its southeastern edge by faults (DRISCOLL and DIEBOLD 1998; MAUFFRET and LEROY 1999). In the north, the contact of Beata Ridge with Hispaniola Island is observed along a transverse NE alignment, which constitutes a morphostructural limit, thus producing the interruption of the Cibao Valley and the divergence of the rivers and basins in opposite directions (COTILLA et al. 2007).

The Beata Ridge is composed mainly of intrusive rocks (gabbros and dolerites) emplaced in a subsurface or hypabyssal environment. The bulk of the ridge most likely represents an imbricated sill/dike complex (RÉVILLON et al. 2000).

The seismicity in the area of Hispaniola Island is mainly distributed throughout the eastern side of Dominican Republic, as well as along the main strike-slip faults (SFZ and EPGFZ); it is marked by a clear discontinuity with the western part, which coincides with the direction of Beata Ridge indentation. Only few shallow earthquakes have been recently located in the northern part of this structure (see INTERNATIONAL SEISMOLOGICAL CENTRE, ON-LINE BULLETIN 2001), but in the past there have been some large earthquakes, such as the one that took place in 1673 with a magnitude of Ms = 7.5. This event was located at the junction of Beata Ridge, Muertos Trough and EPGFZ (COTILLA et al. 2007) (Fig. 1a).

Seismic refraction studies (EWING et al. 1960; EDGAR et al. 1970) show that the crustal structure of the ridge consists of a 5.4 km/s layer overlying a 6.7 km/s layer. The velocity structure is similar to ocean basin velocity structure; however, the total thickness of the Beata Ridge crust is twice as thick as that of normal oceanic basin (FOX et al. 1970). This structure is characterized by unusually thick oceanic crust (up to 20 km) and it is believed to form part of the ancient Caribbean oceanic plateau. Such thickness is mainly caused by a volcanic underplating, which initiated the uplift and rifting of the Beata Ridge (MAUFFRET and LEROY 1999).

In the Caribbean Sea, the Beata Ridge shows a negative magnetic anomaly, while its flanks have positive anomalies. The Venezuelan Basin basically features NE–SW-oriented positive and negative anomalies, subparallel to the Beata Ridge (GHOSH et al. 1984). A recent study shows a set of linear

north–southward magnetic anomalies (east of the Beata Ridge) can be found distributed over a negative terrace (CATALÁN and MARTÍN DÁVILA 2013). This negative horizon is clearly constrained: in the west, by the Beata Ridge; in the north, by a new low-amplitude east–west anomaly that is identified as the Muertos Trough; and as its easternmost limit a negative north–south anomaly that suggests that the magnetic rocks causing these anomalies possesses a strong remanent.

On the CP interior, positive gravity anomalies occur over the western/southern parts of the Beata Ridge and the Lower Nicaragua Rise (JAMES 2009). Moreover, these gravity anomalies show that the Beata Ridge is characterized by strong gradients aligned in a NE–SW direction as the structure is identified in the bathymetry, while the Venezuela Basin shows maximum values of +350 mGal less than NAP in the region of Puerto Rico Trench ([+400 mGal) (GRANJA BRUÑA et al. 2009).

## 3. Data Acquisition and Processing

The CARIBE NORTE project (CARBÓ et al. 2010) was carried out during the spring of 2009 using the Spanish Navy's research ship R/V Hesperides. As a part of this marine experiment, a 220-km wide-angle seismic profile (LM2 line) was surveyed to determine the deep structure across Beata Ridge, between Beata and Alto Velo Islands using wide-angle seismic data. The wide-angle seismic data analyzed in this work were measured along a WNW–ESE trending line (LM2) in the southern coast of Dominican Republic, crossing the northern flank of Beata Ridge, with a total length of 220 km (Fig. 1c). The seismic source used aboard R/V Hesperides consisted of two airgun subarrays with a total capacity of 3850 ci, shooting every 90 s. The cruise velocity was 5 knots. These shots were registered by five OBSs, with the help of the Dominican Navy patrol boat ''Orion'', situated along the LM2 line, and one three-component seismometer LE-3D/lite 1 Hz seismic land station model TAURUS located on Beata Island. The deployed OBSs were short-period model LC2000SP with L28 three-component geophone sensors with a natural frequency of 4.5 Hz and one HiTech HYI-90-

Figure 2
The vertical channel record sections of a OBS 11, b OBS 13, c OBS 14, d OBS 15, e OBS 16 and f BEATA station in seismic line LM2. *Black dashed lines* point out reflected/refracted P-wave horizons. Bathymetry along the *line* is shown on *top* of every seismic record section. Reduction velocity is 6 km/s, the dot was band-pass filtered from 4 to 10 Hz and amplitudes are trace normalized. P-wave phase correlations are represented by $P_X$ = refracted phase in layer number $X$; $P_XP$ = Phase reflected at the bottom of layer number $X$; $P_X^0$ = Head or diving wave traveling across discontinuity between layer number $X$ and $X + 1$

U hydrophone. OBS 11 and 13 were located on the western side of the transect (Haiti sub-basin) and OBS 14, 15 and 16 were to the east, crossing the Beata Ridge. OBS 14 was just below the shooting line, while the others were 1.3 km to 4.0 km offline. The station placed on Beata Island was 13.5 km from the LM2 line. For this reason, it was necessary to apply a correction to the travel times.

The seismic energy was sufficient to trace signals on the OBS record sections to distances up to 220 km. Processing included navigation data, band-pass filtering and corrections due to instrument drift, as well as the previously mentioned travel time correction, except in the case of OBS 14.

Moreover, bathymetry data were collected throughout the CARIBE NORTE cruise aboard the R/V Hespérides, by means the hull-mounted multibeam Kongsberg Simrad EM 120 system. These data have been taken into account to complete the P-wave phase interpretation.

### 3.1. P-wave Phases Determination

In this section, the six seismic records are going to be shown as an example of the work that has been carried out. The first seismic record section that will be analyzed is that of OBS 11. This station was located 4 km above the seafloor, 32 km from the beginning of seismic line and 4 km offline. An error term is needed for the determination of uncertainty, which mainly affects the upper layers of the crust.

Figure 2a shows the seismic record section of OBS 11. The interpretation of the P-wave phases is shown in Fig. 2a. $P_2$ and $P_2P$ phases are observed on western side of the OBS 11 from 3 to 15 km [4.6 km/s average apparent velocity (aav)] and $P_3$ is visible from 8 to 25 km to the east (5.7 km/s aav). On the

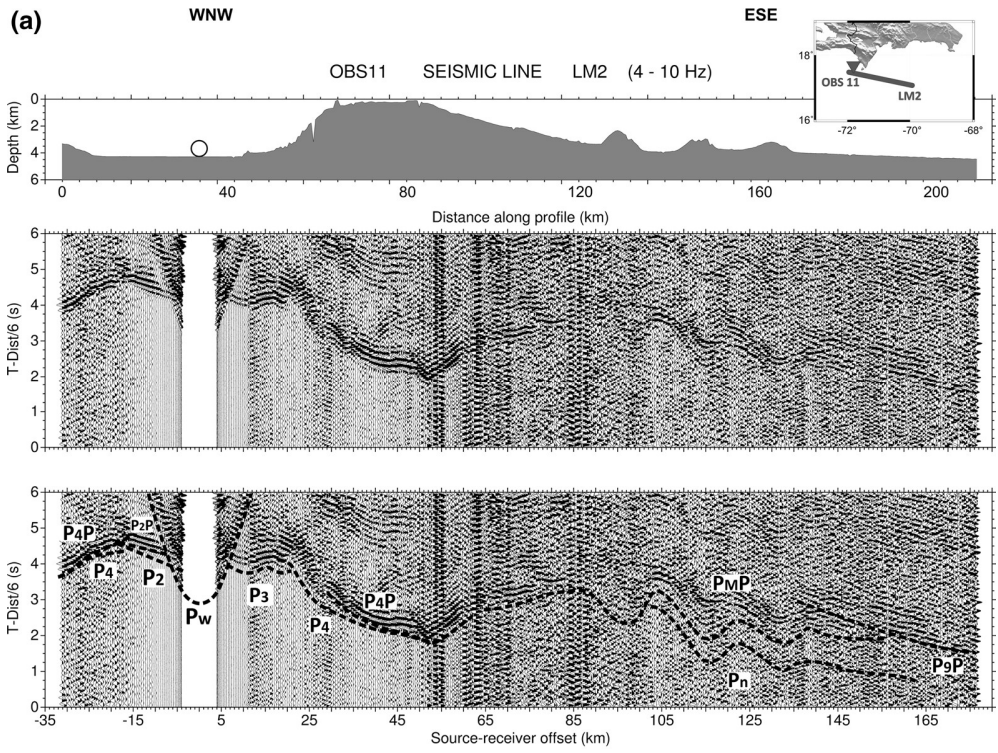
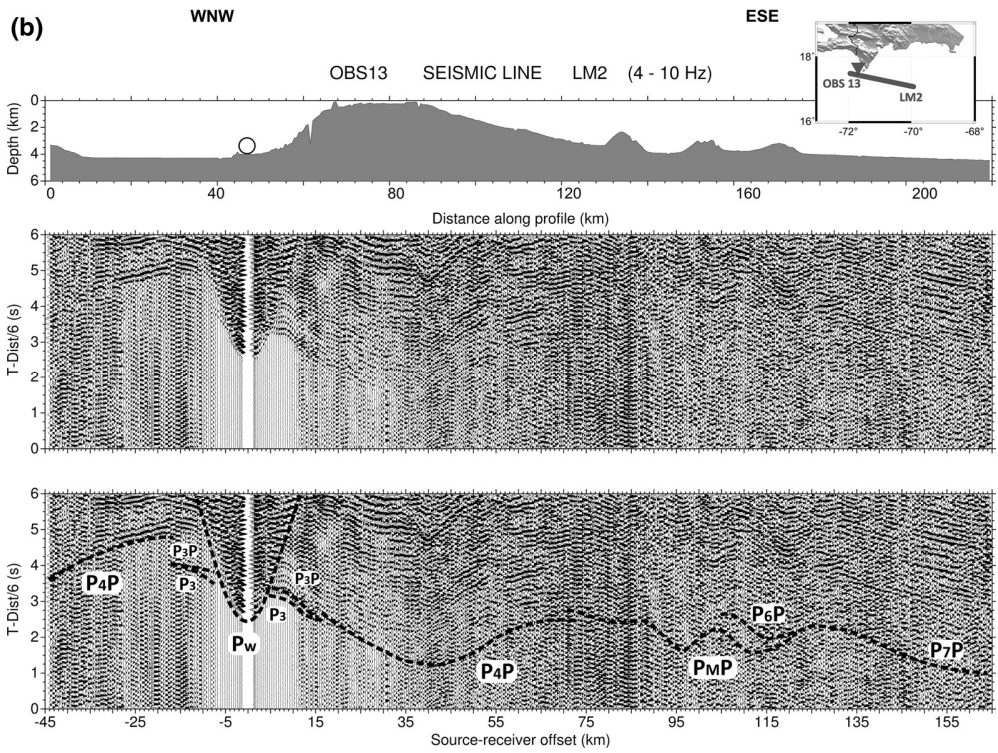

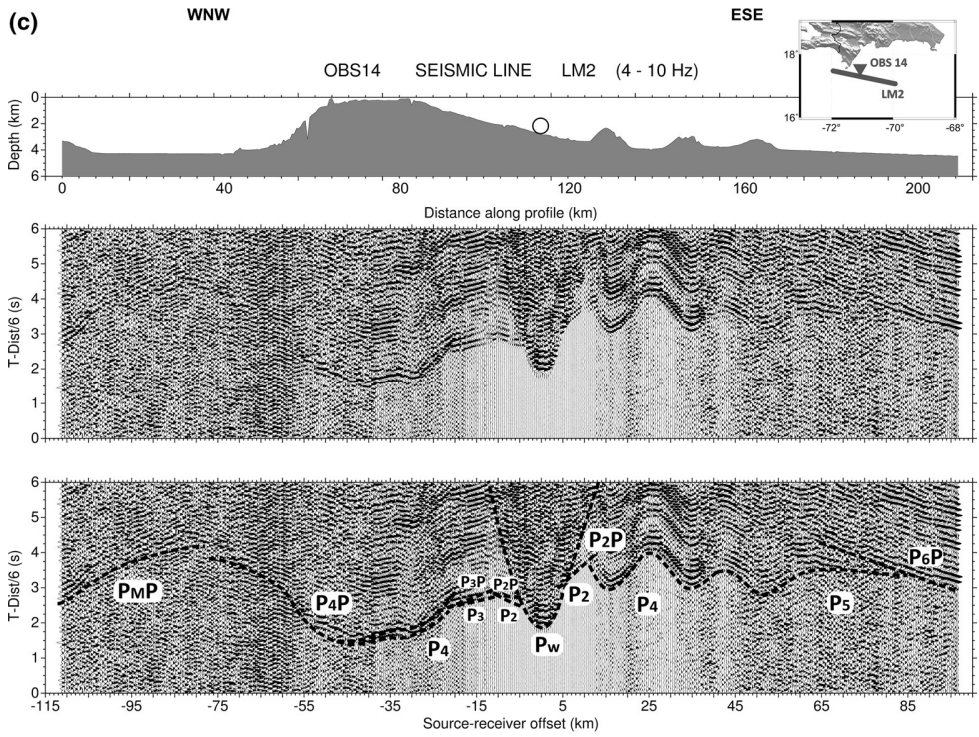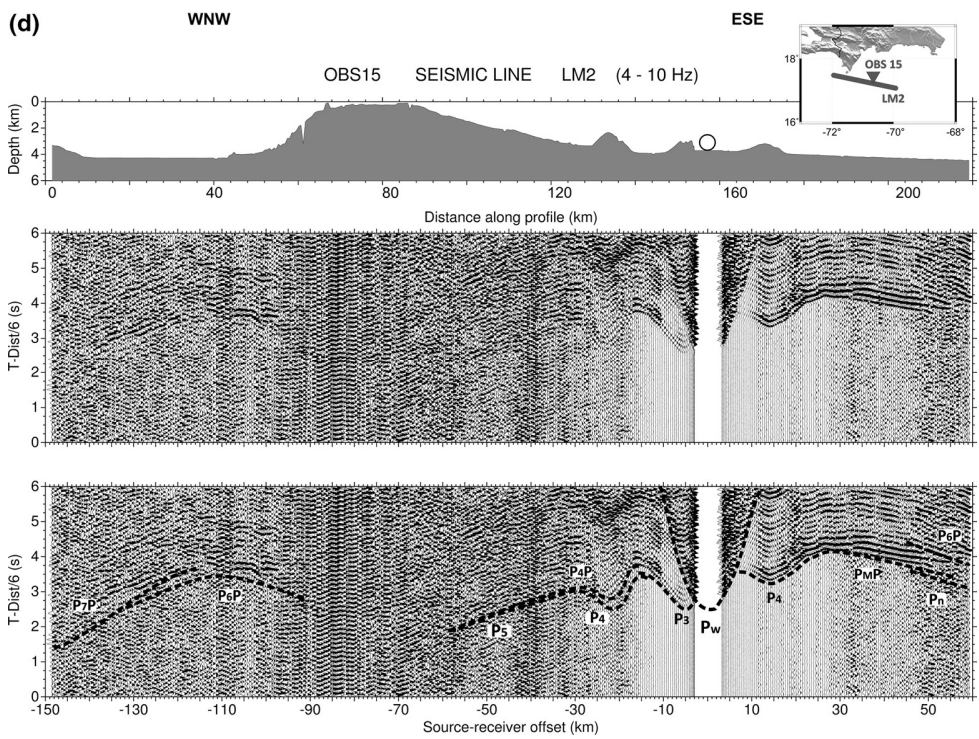

**(e)** WNW — OBS16 SEISMIC LINE LM2 (4 - 10 Hz) — ESE

Phases labeled: $P_6P$, $P_n$, $P_n$, $P_MP$, $P_n$, $P_4'$, $P_4$, $P_2'$, $P_4P$, $P_w$, $P_3$, $P_4'$, $P_2P$

**(f)** WNW — 3C SEISMIC STATION BEATA LM2 (4 - 10 Hz) — ESE

Phases labeled: $P_4'$, $P_3'$, $P_3'$, $P_4'$, $P_4'$, $P_6P$

west, between 17 km to 32 km of source–receiver offset, $P_4$ and $P_4P$ are correlated with an aav of 9.7 km/s, which indicates that layer 4 is dipping.

On the eastern side, it is also possible to observe the $P_3$, $P_4$ and $P_4P$ phases up to 65 km of offset, thus showing correspondence with the bathymetry in the area. The following phase that is observed is the $P_MP$, which can be correlated from 66 to 156 km across Beata Ridge and, also, a $P_n$ phase between 102 and 162 km of distance (7.6 km/s of aav). Deeper phase, $P_9P$ is observed from 148 km to the end of the seismic record section.

To the west of Beata Island, OBS 13 was deployed at 3.9 km below sea level, 13 km to the east of OBS 11 and 1.1 km offline. The P-wave phases of this seismic record section are displayed in Fig. 2b. The first phases identified correspond to $P_3$ and $P_3P$ from 5 to 13 km of offset and an aav of 3.8 km/s for refracted phase. Then, it is interpreted $P_4P$ correlated between 17 and 44 km offset. In the eastern side of the record section (Fig. 2b), the P-wave phases identified are: $P_3$, $P_3P$, $P_4P$, $P_MP$, $P_6P$ and $P_7P$. First phase corresponds to a refracted phase in third seismic layer from 5 to 15 km and 6.6 km/s of aav, indicating that this layer is dipping. The following phase is the corresponding reflected phase at the same offset. The $P_4P$ phase is identified between 15 and 75 km of offset and followed by the $P_MP$, which is observed in an offset range between 75 and 128 km. The deepest phases interpreted are the reflected phases over sixth and seventh layers from 100 to 130 km and 125 to 165 km of offset, respectively.

To the east of Beata Island, OBS 14 was deployed at 112 km distance from the beginning of seismic line LM2 and 2.8 km below sea level. This OBS is located in the marine shooting line; therefore, it is not necessary to add any additional error term. The P-wave phases identified on the western side of this seismic record section (Fig. 2c) correspond to $P_2$, $P_2P$, $P_3$, $P_3P$, $P_4$, $P_4P$ and $P_MP$. The first phases correspond to a refracted waves from the second layer and reflected waves at the discontinuity between second and third layers, which can be followed from 3 to 10 km offset range with an aav of 4.9 km/s. The refracted and reflected phases of the third layer can be correlated between 10 and 22 km offset and 7.6 km/s of aav. The phases correlated in surrounding area of phases have been interpreted, $P_3$, $P_2P$ and $P_4$. The first one is visible from 5 to 12 km, second one from 4 to 16 km and the last one from 12 to 24 km offset. The average apparent velocity obtained for $P_4$ phase is

Beata Island are $P_4$ and $P_4P$, which are observed from 20 to 75 km of distance, followed by $P_MP$, which is correlated up to the end of the seismic record section (112 km offset).

To the east of OBS 14 seismic record section describes the three seamounts that characterize this area. Between 5 and 15 km, $P_2$ and $P_2P$ phases are correlated with an aav of 3.9 km/s, followed by $P_4$ and $P_5$ with 10–40 and 40–85 km of respective offset ranges. Last phase identified in this seismic record section is the phase reflected in the sixth layer, $P_6P$, which is between 60 km and 97 km.

The seismic record section of OBS 15 is shown in Fig. 2d. This OBS is located at 148 km from the beginning of the seismic line LM2 and 3.3 km below the sea level. For the upper layers, an error term has been included to the uncertainty as this station is 3 km offline. The first identified P-wave phases in this figure show correspondence with the bathymetry on the both sides of the OBS. Towards to Beata Island, the first phases are $P_3$, $P_4$ and $P_4P$ correlated between 4 and 20 and from 10 to 39 km of offset, respectively, with aav of 4.2 and 6.6 km/s. To the west, these phases are followed by $P_5$ in an offset range of 40–55 km with an aav of 8.4 km/s. The reflected phase $P_6P$ is correlated from 89 to 133 km of distance, while the deepest phase corresponding to $P_7P$ is visible between 112 and 146 km source–receiver offset in the area of Dominican Sub-basin.

To the east of OBS 15, the seismic record section shows a refracted wave from the fourth layer, $P_4$, which is correlated with an aav of 5.1 km/s between 6 and 26 km. The following phases that are visible from 26 to 54 km and 48 to 59 km offset correspond to $P_MP$ and $P_n$ phases. Deeper phase, $P_6P$ is correlated up to the end of the seismic record Section (59 km of offset).

Figure 2e shows the seismic record section of OBS 16. This OBS is the last station located in the profile at 185 and 4.2 km below sea level. This station is 3.5 km offline; therefore, an error term is added to the uncertainty for upper layers. In the study to estimate this error, it was found that the separation between the station and the line mainly affects the first three layers and phases that are closest to the station.

The eastern part of the record section is shorter than that of the previous section; therefore, three close to 3.7 km/s. West of the OBS 16, $P_2^0$ is found from 5 to 17 km distance with 7.3 km/s of aav, thus indicating a dip due to the presence of a third seamount of Beata Ridge. After this phase, $P_3$ is correlated from 5 to 16 km offset range and then, $P_4$ and $P_4P$ are correlated between 17 and 35 km offset. These phases are followed by the $P_MP$ with 35 to 65 km offset. The diving wave that travels through the Moho discontinuity, $P_n$, is visible from 65 to 148 km source–receiver offset in the area of Beata Island. In this profile, $P_6P$ is correlated between 148 and 185 km offset in the area of Haiti Basin.

Moreover, in the OBS 16 seismic record section it can be observed that the amplitudes of multiple waves are stronger than in the other seismic record sections of this profile.

The portable seismic station BEATA is located on Beata Island, 6 km south of the Beata Cape (DR). This station lies 13.5 km offline; therefore, the estimated error is much higher than that of the others stations, reaching values greater than 1 s for the first layers.

In the seismic record section that corresponds to BEATA seismic station (Fig. 2f), high absorption in the seismic signal in the western and eastern flanks has been observed. Towards Haiti Basin, $P_3^0$ is visible from 14 to 38 km offset, followed by the head wave which travels through the fourth layer discontinuity, $P_4^0$, between 38 and 78 km. The eastern part shows the correlation of $P_3^0$, $P_4^0$ and $P_6P$. The $P_3^0$, whose shape, in correspondence with the bathymetry, is correlated from 14 to 48 km, while the $P_4^0$ is correlated from 60 to 108 km source–receiver offset. At longer distances the correlation is not clear due to noise. Finally, $P_6P$ is correlated between 78 and 132 km offset range.

## 4. Uncertainty Estimation

### 4.1. Uncertainties in Arrival Time Picking

After data processing, phase determination and phase picking, we have estimated uncertainties for every phase due to two main error sources: (1) arrival time picking, and (2) offset between seismometer position and seismic line. The uncertainties in arrival time picking were determined using the methodology developed for seismograms by DIEHL et al. (2009) and applied to the seismic record sections obtained in our experiment (NÚÑEZ et al. 2011).

The criterion used in this study is in accordance with the criterion proposed by DIEHL et al. (2009). However, the strength of refraction seismology was included: we can establish phase correlations along seismic arrays. This fact means that every pick can be determined with its appropriated uncertainty relating trace-to-trace and, also, obtaining correlation uncertainty.

The first error calculation considers the onset of a seismic phase as a probabilistic function $P_a(t)$, the arrival time is expressed as the "most likely" time $t_A$, with $P_a(t_A) = Max(P_a)$. The "earliest" possible time for the phase onset is defined as $t_E$. Similarly, $t_L$ is defined as the "latest" possible time for the phase onset (DIEHL et al. 2009).

In practice, $t_L$ has been determined as the interception between the signal amplitude and the a priori noise threshold, which is defined as 1.5 times the noise amplitude. Meanwhile, $t_E$ is defined as the first zero slope of a tangent from $t_L$. Subsequently, the arrival of the phase is picked at the most likely position $t_A$, between the error interval of $t_E$ and $t_L$.

Hence, our methodology consists of (1) picking each wavelet separately using the principles for single signal picking according to DIEHL et al. (2009); (2) defining the correlation of phases by following the principles of refraction seismology, and then, (3) repicking the earliest phase onset time for each trace, while taking into account the entire phase and not just the single trace.

In our study, we have seismic record sections that correspond to one seismic station that records a line of shots. Then, due to the dense number of traces whose phases can be followed by a distance interval, sometimes, it is not possible to extract single trace travel times. In those cases, the error estimation is calculated for every phase as an average of the individual error estimations.

## 4.2. Offset Between the Seismometer Position and the Seismic Line

The simplest case corresponds to a two-dimensional reflection at a horizontal boundary. In this case, the ray that strikes the boundary at $R$ is reflected to the surface and recorded by a geophone at the point G (Fig. 3). When the seismometer is away from the seismic line with an offset $x_S$, the new travel time, $t'$, corresponds to the reflected ray SR'G (Eq. 1).

$$t' = \frac{SR' + R'G}{V} = \frac{\sqrt{x'^2 + (2d)^2}}{V} \quad (1)$$

where

$$x' = \sqrt{x^2 + x_S^2} \quad (2)$$

The delay in travel time is considerable and calculated as:

$$\Delta t = t' - t = \frac{1}{V}\left[\sqrt{x'^2 + (2d)^2} - \sqrt{x^2 + (2d)^2}\right] \quad (3)$$

For a more complex layered model, we calculated the difference between the travel times from a station on and off seismic line by ray tracing. For this purpose, we considered a realistic model (velocity model characterized by bathymetry and successive layers that may have dips) in which absolute travel times do not vary regardless of the seismometer position. The results show that the error is within the phase picking uncertainty for $x_s \leq 1.5$ km. Most of stations are in this offset range, but some were more offline than 1.5 km. In those cases, the error estimation has been added to phase picking uncertainty.

## 4.3. Depth and Velocity Error Estimation

Considering that the possible error sources are travel time ($T$), distance ($x$), depth ($h$) and velocity ($v$), and using a plane layered medium as example, Eq. 4 describes the travel time for a wave refracted along the top of the $n$th layer of one model with $n$ uniform horizontal layers of $h_j$ thickness and P-wave velocity of $v_n$. From this equation, a minimum uncertainty value for depth and velocity is going to be estimated from a general formula of error propagation (Eq. 5), all while assuming some approximations:

$$T_{H_n} = \frac{x}{nv} + 2\sum_{j=0}^{n-1} h_j \left(\frac{1}{v^2} - \frac{1}{v_n^2}\right)^{1/2} \quad (4)$$

$$\Delta T_H = \frac{\partial T_H}{\partial x}\Delta x + \frac{\partial T_H}{\partial h}\Delta h + \frac{\partial T_H}{\partial v}\Delta v \quad (5)$$

Depth: to determine depth uncertainty, $\Delta h$, we assume that $\Delta x = \Delta v = 0$, this uncertainty, therefore, only depends on absolute travel time, whose value has been determined in the previous sections. In this case,

$$\frac{\partial T_H}{\partial h} \quad (6)$$

Using the data of velocity model, it is possible to observe that minimum depth uncertainty increases as the absolute travel time uncertainty increases. In this case, depth uncertainty varies from 0.03 to 0.4 km for the shallowest part and in a range of 1–2 km for the deepest layers in the applied model.

Velocity: to calculate velocity uncertainty, $\Delta v$, we assume that $\Delta x = \Delta h = 0$. In this case, velocity uncertainty only depends on the absolute travel time, whose value has been determined in the previous sections. Using the same data as before, the minimum

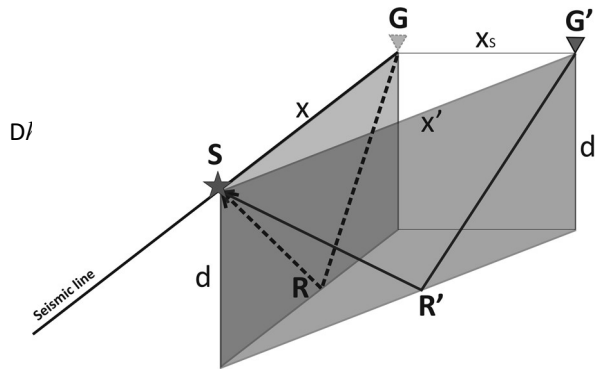

Figure 3
Two-dimensional reflection scheme representing offset between geophone position and seismic profile. $S$ Shot point, $G$ receiver point (geophone) over seismic line, $G'$ receiver point offline, $R$ point where reflection occurs in the seismic line, $R'$ point where reflection occurs offline. The distance between G and S is represented by $x'$ and separation distance is $x_S$. Depth of the layer is $d$ velocity uncertainty increases in distance and depth, obtaining maximum values between 0.05 and 0.1 km/s for the shallowest part and up to 0.3 km/s for the deepest layers.

$$\Delta v = \frac{\Delta T_H}{\frac{\partial T_H}{\partial v}} \quad (7)$$

This new methodology allows us to obtain minimum values of velocity and depth uncertainties for our crustal velocity models, due to travel time uncertainty. Moreover, the relative position between the seismometer and the seismic line has been considered, thus quantifying the error.

## 5. Modeling

The interpretation of crustal refraction data for a 2D velocity structure often involves laborious trial-and-error ray-trace forward modeling. The theoretical travel time and amplitude response of a laterally inhomogeneous medium are repeatedly compared with observed seismic record sections, a model which provides a satisfactory match between calculation and observation is found (ZELT and ELLIS 1988). During this study, forward modeling and travel time inversion have been used to obtain 2D velocity and interface structure. In modeling, we used programs developed by ZELT and SMITH (1992).

This method is applicable to any set of travel times for which forward modeling is possible, regardless of the shot-receiver geometry or data quality, since the forward step is equivalent to trial-and-error forward modeling (ZELT and SMITH 1992). This methodology requires specifying the number and position of the velocity and boundary nodes for each layer. The method of ray tracing is carried out using an efficient numerical solution of the 2D ray tracing-coupled equations with an automatic determination of ray take-off angles.

$$\frac{dx}{dz} = \tan\theta \quad (8)$$

$$\frac{d\theta}{dz} = \frac{v_z \tan\theta - v_x}{v} \quad (9)$$

The initial conditions are: $x = x_0$, $z = z_0$, $\theta = \theta_0$ (CERVENY et al. 1977). To complete the basic ray tracing algorithm, Snell's law is applied at the intersection of a ray with a layer boundary.

The effectiveness of this inverse technique is due to a model parameterization that is suited to the requirements of an inverse approach, where the number and position of velocity and boundary nodes can be suited to the data subsurface ray coverage. The forward step (TRAMP) uses a robust method of ray tracing and, therefore, the inversion algorithm benefits from the advantages of ray methods (RAYINVR). A simulation of smooth layer boundaries increases the stability of the inversion. Synthetic seismograms have been calculated using PLTSYN program included in Zelt's software package (ZELT and SMITH 1992).

## 6. P-wave Model Interpretation

Data processing has allowed us to obtain record sections whose interpretation has provided the travel times of refracted and reflected P-waves and distances necessary to elaborate the northern Beata Ridge velocity model, which will be presented in Fig. 10. In this section, we have divided the model into shallow, middle and deep parts, to analyze to the greatest extent possible. This interpretation consisted of determining these P-wave phases observed in each seismic record by the correlation process and their corresponding apparent velocities providing the basis for the determination of velocity distribution and depths of seismic boundaries in the crust and uppermost mantle. Subsequently, ray tracing models and synthetic seismograms have been computed to produce the final model that fits the data. The seismic wide-angle data have been interpreted using forward modeling techniques (ZELT and SMITH 1992) as indicated in the previous section.

The proposed velocity model is defined in terms of seismic layers separated by discontinuities at specified depths defined by user. Moreover, the values of velocity and depth have been adjusted manually in forward modeling step and by damped least squares inversion.

The final velocity model has 7 layers and 2 floating reflectors constrained by the data by reflections in velocity discontinuities. This model reproduced a total of 1711 of the 1785 observed travel times. Water depth/elevation from the airgun-shots/land stations was taken from the navigation and bathymetry data provided by R/V Hespérides. We will hereafter refer to model distance along this transect, which is measured from a point located 32 km northwest of OBS 11.

### 6.1. Shallow Crustal Structure

On the western part of Beata Island (Haiti Basin), water depth reaches maximum values of 4.3 km of depth between 20 and 30 km of model distance. This layer produces a substantial influence on travel times with maximum delays close to 3 s. In this area, the sedimentary cover has two layers, whose depth has been defined using the data from OBS 11 and OBS 13 (Fig. 4). The thickness of first sedimentary layer is 2.3 km, which decreases towards Beata Island to 1.2 km, with an estimated uncertainty less than 1 km. In this layer, the P-wave velocity varies between 3.3 and 3.8 ± 0.1 km/s. The layer that is located immediately below it reaches 4.6 km depth with a maximum $V_P$ of 4.8 ± 0.1 km/s. This layer, like the previous one, rises up near Beata Island (Fig. 4).

The area to the east of Beata Island is characterized by the influence of Beata Ridge shown as three seamounts in the model. The structure can be associated with geodynamic blocks in the eastern Hispaniola proposed by COTILLA *et al.* (2007). The shallow structure (Fig. 5) is determined by the wide-angle seismic data of OBS 14, OBS 15 and OBS 16. At the model distance, 90–120 km, water depth elevation increases up to 3.3 km and the sedimentary cover thickens, reaching a maximum value of 2.2 km for the first layer and 2.1 km for the second one. These last layers are strongly influenced by the bathymetry of the area. The first layer disappears near

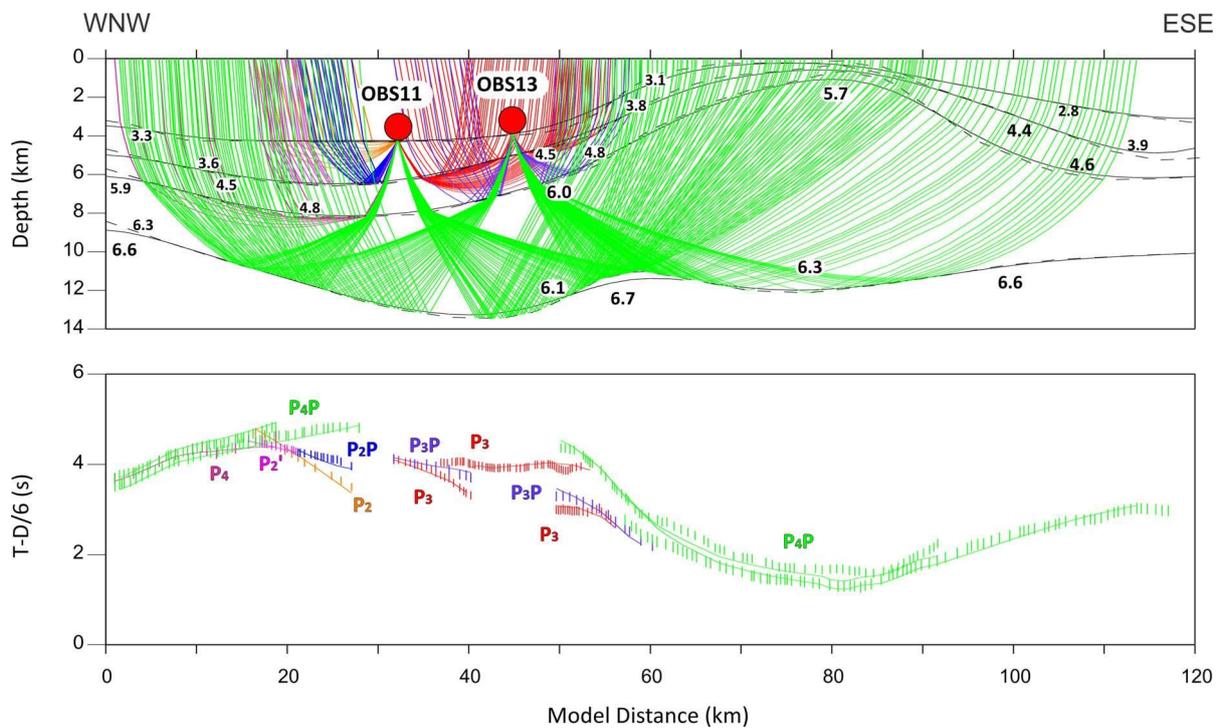

Figure 4

*Top* Ray tracing corresponding to shallow structure in Haiti Basin and Beata Island with OBS 11 and OBS 13 marine stations and velocity model with velocities in km/s. *Dashed lines* overlapped *continuous lines* represent connections between the velocity and boundary nodes specified for each layer in the P-wave velocity model. *Bottom* Comparison between calculated (*lines*) and observed (*vertical bars*) travel times whose height is the uncertainty estimated for every phase. Distances refer to the origin of the velocity model. *Red circles* represent the location of the stations in the model

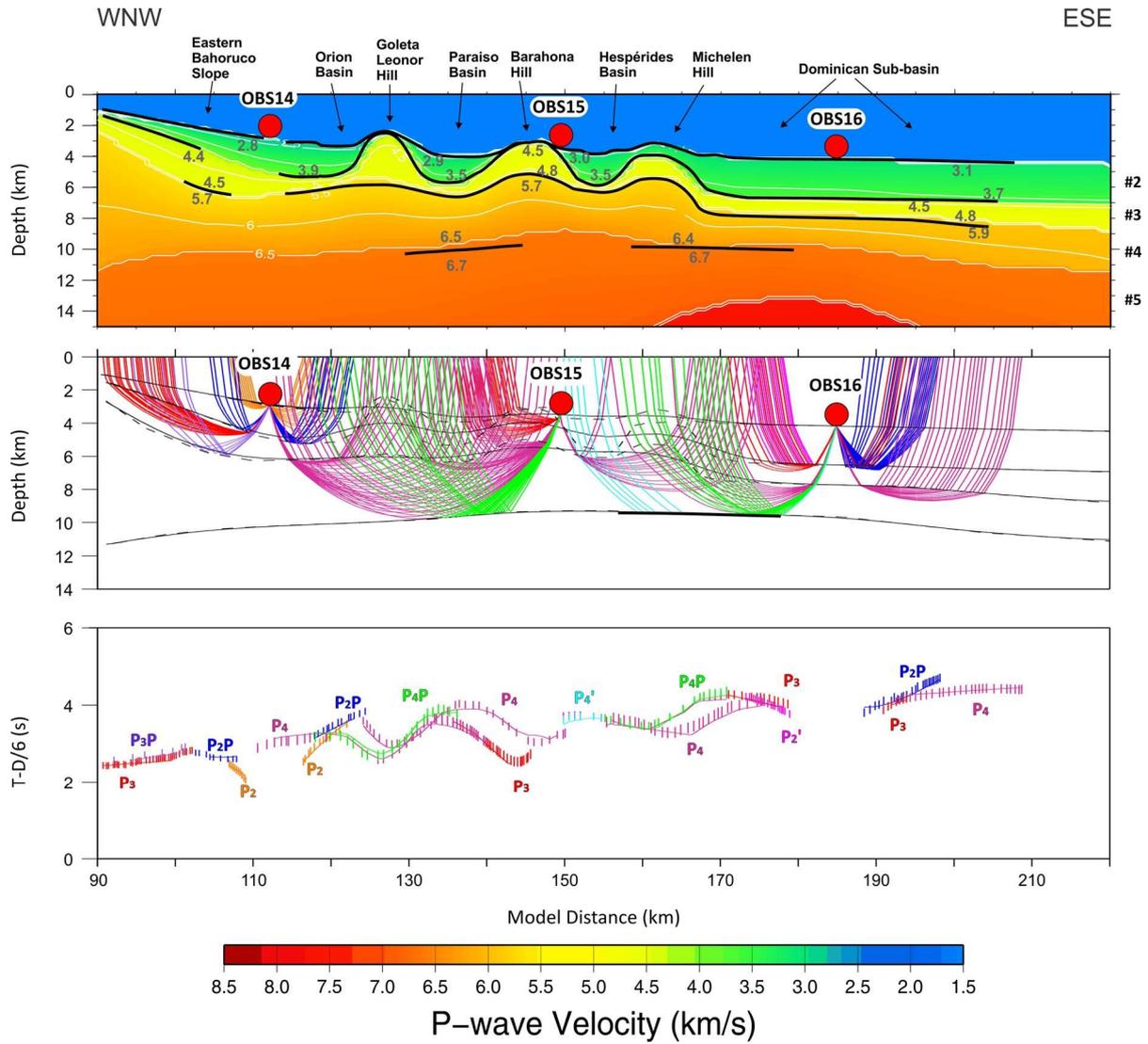

Figure 5

*Top* P-wave velocity model with velocity contours every 0.5 km/s in the area of Beata Ridge. *Red circles* show the seismic stations used in this area. *Horizontal axis* shows model position in km and vertical scale corresponds to the depth below the surface. The *colored area* is the area traversed by refracted rays, indicating the area where velocities are constrained. *Thick black lines* mark positions where rays are reflected and refracted, showing well-constrained boundary positions. The layer number is represented by #X, where X is the layer number. Color scale denotes P-wave velocity in km/s. *Middle* Ray tracing corresponding to shallow structure in Beata Ridge with OBS 14, OBS 15 and OBS 16 marked. *Dashed lines* overlapped *continuous lines* represent connections between the velocity and boundary nodes specified for each layer in the P-wave velocity model. *Bottom* Comparison between calculated (*lines*) and observed (*vertical bars*) travel times whose height is the uncertainty estimated for every phase. Distances refer to the origin of the velocity model the top of the seamounts and the vicinity of Beata Island. The second layer, however, is practically constant in thickness, reaching up to 6.4 km below westernmost seamount. Between the western and middle seamounts, the first sedimentary layer is 2.0 km thick and the second one is 1.2 km thick. The thicknesses between middle and eastern seamounts are 1.8 and 1.5 km, respectively, remaining practically constant in the Venezuela Basin area where sea floor is located at 4 km depth.

When comparing the structure that has been determined by this paper with a previous MCS study

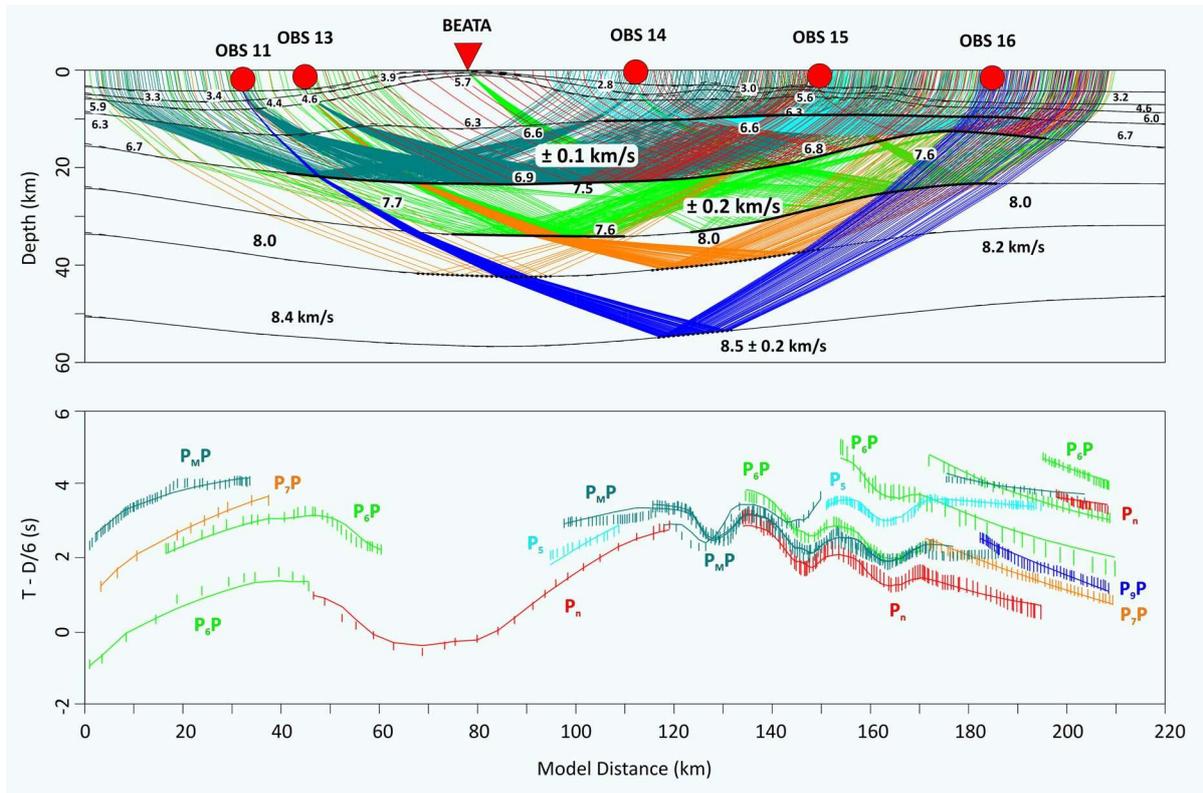

Figure 6
*Top* Ray tracing corresponding to CARIBE NORTE seismic line across Beata Ridge and velocity model with average velocities in the corresponding layer in km/s. *Dashed lines* overlapped *continuous lines* represent connections between the velocity and boundary nodes specified for each layer in the P-wave velocity model. The *red circles* and *inverse triangle* represent the seismic stations used in this study (OBS 11, OBS 13, OBS 14, OBS 15, OBS 16 and BEATA land station). *Bottom* Comparison between calculated (*lines*) and observed (*vertical bars*) travel times whose height is the uncertainty estimated for every phase. Distances refer to the origin of the velocity model

(Mauffret and Leroy 1999), it is possible to establish a correlation between the first sedimentary layer and Horizon A″, while second sedimentary layer corresponds to Horizon B″. Mauffret and Leroy (1999) define Horizon A″ and B″ as reflectors that form the base of Middle to Early Miocene chalks and Santonian to Conician basalts intervals, respectively. Both reflectors characterize the seismic stratigraphy of Venezuela Basin. These correlations allow us to determine the composition of the sedimentary cover. As was mentioned before, the first layer of this study corresponds to Middle Eocene to Early Miocene chalks, while the second one corresponds to the Santonian to Conician basalts. In this way, the thinning of the A″–B″ layer (Fig. 6) and the old sedimentary fill in the study area suggest that these features are contemporaneous and are related to volcanic plateau formation (Mauffret and Leroy 1999). The estimation of depth uncertainty provides values of less than 1 km in the shallow part of the model.

### 6.2. Middle and Lower Crustal Structures, Moho Boundary and Upper Mantle

The middle crust is 4 km thick to the west of Beata Island and decreases in depth but not in thickness between 30 and 100 km of model distance, where it reaches a maximum thickness of 9 km with a $V_P$ value range of 5.7–6.3 ± 0.1 km/s. Below the island, the middle crust thickness remains practically constant at 9–10 km up to Venezuela Basin area.

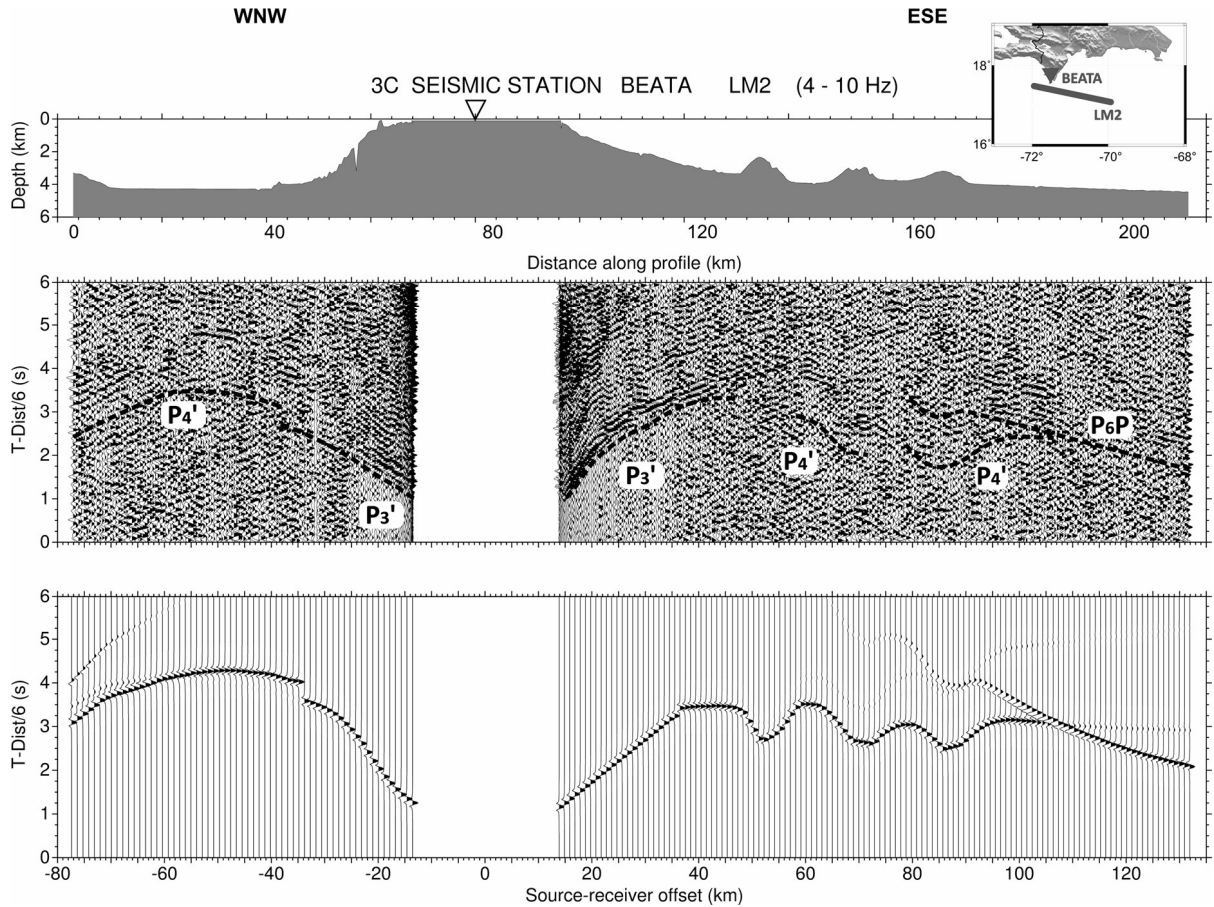

Figure 7
*Top* Bathymetric profile. *Middle* Vertical seismic section, correlated seismic phases are marked with *dashed line*. *Bottom* Synthetic seismogram corresponding to land station BEATA recording marine line LM2. Amplitudes are trace normalized

The Moho depth is 20 ± 2 km in the west, and it increases to 24 ± 2 km below Beata Island. The P-wave velocity is 6.7 ± 0.1 km/s on average in lower crust and 7.6 ± 0.2 km/s below the Moho (Fig. 6). In this part of the model, $P_6P$ and $P_7P$ determine one deeper layer and one floating reflector with $V_P$ of 8.0 and 8.3 km/s, respectively, and with an uncertainty of 0.2 km/s, reaching a maximum depth of 42 ± 3 km. These values are not absolutely claimed because we do not have seismic rays in both directions. Below Beata Island (Fig. 7), Moho depth reaches 24 ± 2 km and P-wave velocity contrast at Moho boundary is 6.6 to 7.7 km/s with 0.2 km/s of estimated uncertainty.

Figure 8 shows ray tracing diagram, observed and computed travel times and uncertainty for BEATA land station. The uncertainty stems mainly from handpicking error of the onsets (NUÑEZ *et al.* 2011) and from the fact that the station is 13.5 km away from the seismic line.

In eastern Beata Island region (Fig. 9), it is possible to find Moho discontinuity at 24 km of depth, rising up to 13 km of depth at the model distance of 170 km (Fig. 6). The next layer is located at 34 ± 2 km depth and rises, following Moho boundary, up to 23 km with a $V_P$ contrast of 7.6 and 8.0 ± 0.2 km/s, which are determined by several rays reflecting in this layer (Fig. 6). The study of relative amplitudes reveals that observed and theoretical data are comparable since they are within the uncertainty interval (Figs. 7, 9). The deepest floating reflector interpreted from wide-

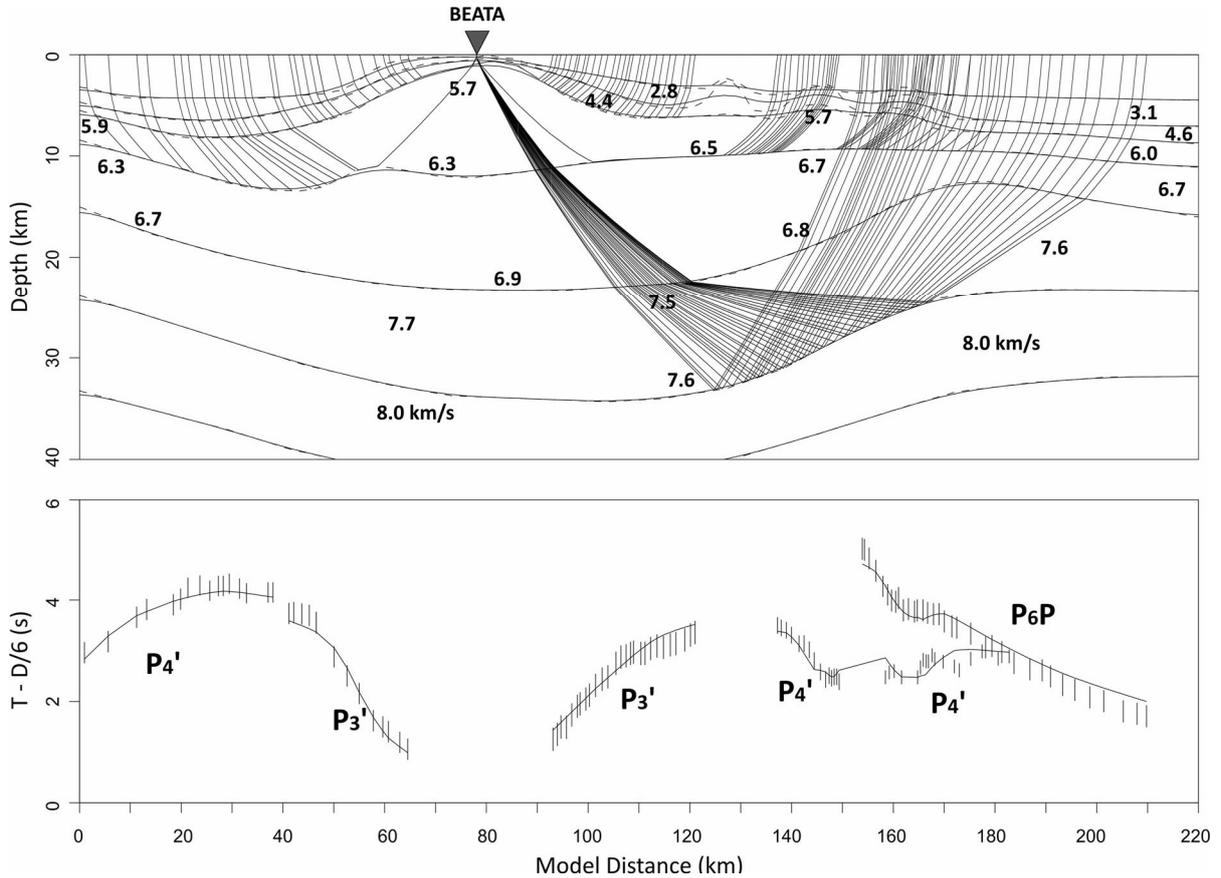

Figure 8
*Top* Ray tracing corresponding to BEATA land station and velocity model with average velocities in the corresponding layer in km/s. *Dash lines* represent boundaries between layers proposed by these authors, while s*olid black lines* represent smoothed boundaries used by forward and inversion programs. *Bottom* Comparison between calculated (*lines*) and observed (*vertical bars*) travel times whose height is the uncertainty estimated for every phase. Distances refer to the origin of the velocity model. The *inverse triangle* represents the location of the station in the model angle seismic data of this line is found from 125 to 150 km with a depth between 50 and 55 ± 3 km. Below this reflector, P-wave velocity is 8.5 ± 0.2 km/s (Fig. 10).

East of Beata Island, it is possible to observe a variation of seismic energy, especially at the Beata Island record section (Figs. 2f, 7), which could be associated with a more complicated structure. The OBS record sections located to the west of the island do not show similar variation of energy. The OBS located to the east of Beata Island shows less energy. This effect is more pronounced in the easternmost seamount.

### 6.3. Travel Time Fit and Model Evaluation

A final velocity model should adequately fit the data predicting arrival times within the data error bounds, ideally with $v^2 = 1$ (AFILHADO *et al*. 2008). If $v^2 \backslash 1$ the data are over-fit and if $v^2 [ 1$ the data are under-fit (ZELT and SMITH 1992). However, in practice, the final $v^2$ values that are significantly different from 1 are often obtained through travel time inversion (ZELT and FORSYTH 1994; ZELT 1999). Estimates of arrival time fit quality for each phase and for all phases are given in Table 1. Our final model produces a normalized $v^2$ of 0.93, close to the ideal case, with the most of individual phases over-fit ($v^2\backslash 1$).

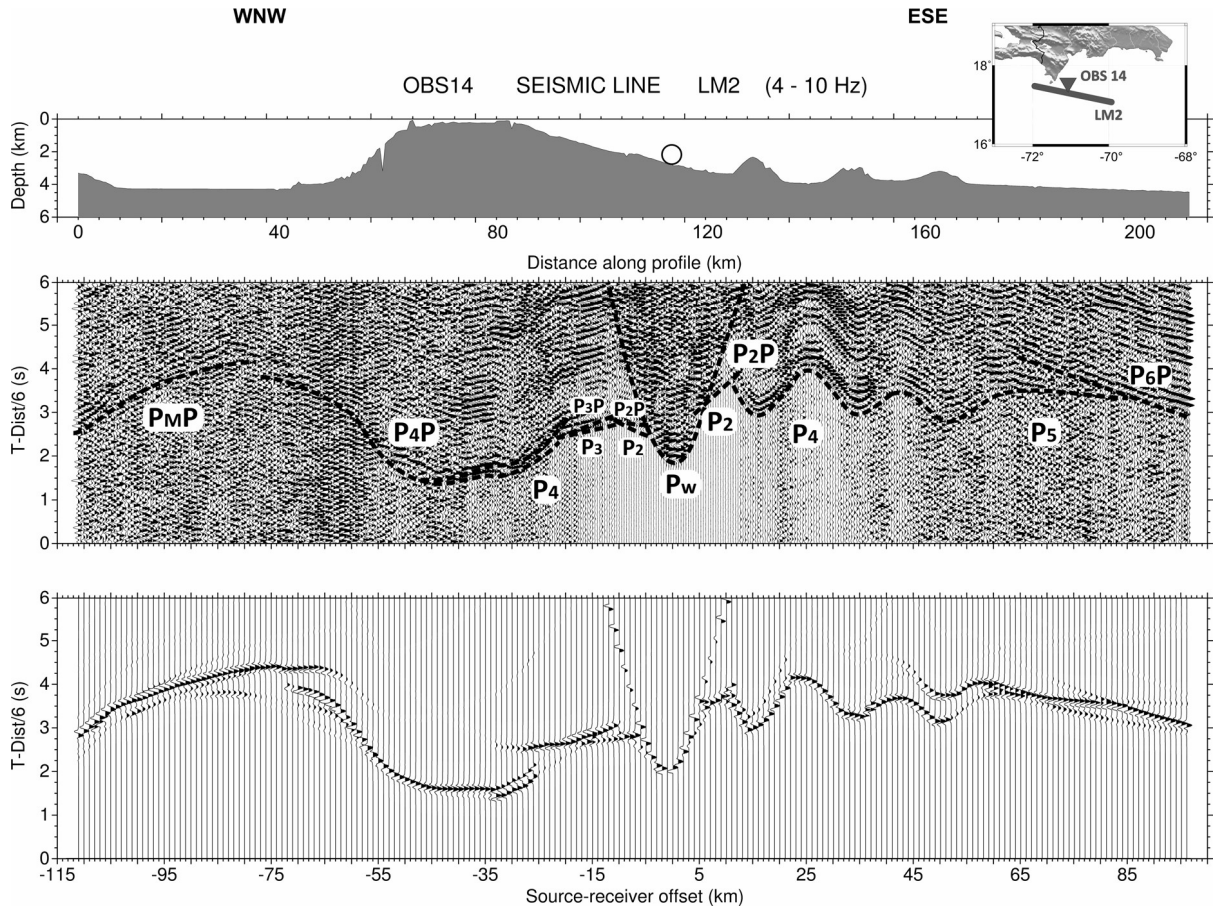

Figure 9

*Top* Bathymetric profile. *Middle* Vertical seismic section with correlated phases. Amplitudes are trace normalized. *Bottom* Synthetic seismogram corresponding to OBS 14 recording on marine line LM2. Amplitudes are trace normalized

Additionally, we have checked the accuracy of the Moho boundary depth from different lower crust velocities (6.7, 7.0 and 7.2 km/s). The results (Table 2) show that in the areas of Haiti sub-basin and Beata Ridge, the Moho experiments lower variations in the depth than in Beata Island region. Maximum differences are more than 3 km with respect to the proposed model in this paper and correspond to a lower crust velocity of 7.2 km/s.

## 7. Discussion

The analysis of wide-angle seismic data for LM2 profile obtained in CARIBE NORTE project (2009) reveals new features about the northern part of Beata Ridge, which collides with the central part of the Hispaniola. In this section, a summary of the velocity model is presented and compared with all other studies carried out in the study area.

The proposed model presents tectonic differences between the western and eastern zones that are separated by Beata Island. In the western area, two sedimentary layers are found. The top of this layer has a $V_P$ of 3.3–3.6 ± 0.1 km/s and the second one shows a maximum velocity of 4.5–4.8 ± 0.1 km/s, characterizing the sequence with continuous highly reflective reflectors that correspond with ponded sediments, which are present in Haiti sub-basin (GRANJA-BRUÑA *et al.* 2014). The maximum total thickness is 4.4 km, as observed below OBS 11. These layers practically disappear (*0.4 km of

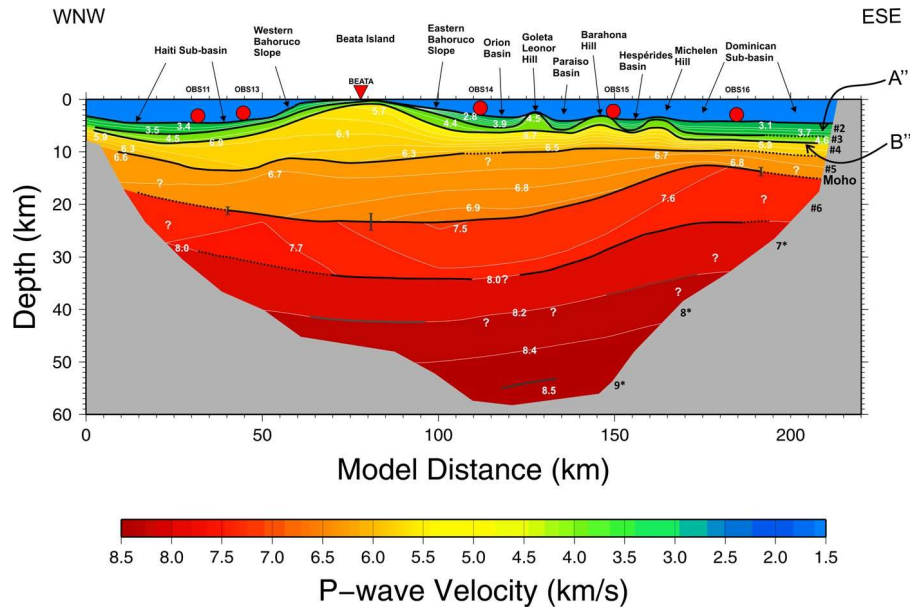

Figure 10
P-wave velocity model of the CARIBE NORTE seismic profile across northern flank of Beata Ridge. *Red circles* and *inverted triangle* show the seismic stations used in this study, from west to east are: OBS 11, OBS 13, BEATA, OBS 14, OBS 15 and OBS 16. *Horizontal axis* shows model position in km and *vertical scale* corresponds to the depth below the surface. The *colored area* is the area traversed by refracted rays, indicating the area where velocities are constrained. Layer boundaries are described by *black lines* and *thick lines* mark positions where rays are reflected, showing well-constrained boundary positions. *Grey thick lines* represent reflections over floating reflectors, showing not well-constrained boundary positions. *Vertical bars* represent depth uncertainty obtained for Moho discontinuity. *Grey zone* denotes the area not crossed by rays and *white question marks* represent regions not well-controlled by ray coverage. The layer number is represented by #X, where X is the layer number, while X* represents floating reflectors. Figure is scaled according depth and distance 1.5:1. Color scale denotes P-wave velocity in km/s. A″ and B″ represent the horizons proposed by Mauffrett and Leroy (1997) that characterize reflectors of Venezuela Basin related to Middle to Early Miocene chalks and Santonian to Conician basalts intervals, respectively thickness) in the vicinity of Beata Island (Western Bahoruco Slope), and then reappear between the three seamounts or hills (*1.5 km thick) that characterize the northern Beata Ridge region, filling in Orion, Paraíso and Hespérides Basins with sediments and accumulations of blocks derived from slope failures in the eastern steep slope off Bahoruco peninsula (Granja-Bruña *et al.* 2014) (Fig. 10). From west to east, these seamounts are: Goleta Leonor, Barahona and Michelen Hills, characterized by an estimated width of 12, 14 and 13 km and a height of 2.9, 2.3 and 2.6 km, respectively. Those seamounts have linear geometry in a map view and triangular shape in cross-section view. They are interpreted as igneous constructions controlled by faulting (Mauffret and Leroy 1997; Diebold 2009). According to Mauffret and Leroy (1999), two of these seamounts have conical symmetry and the other one is asymmetrical. We observed the loss of seismic energy in the record sections corresponding to the OBS presented in this paper. From a qualitative point of view, we have plotted the azimuthal record sections at Profile A (station A1 western Beata Ridge) and Profile B (station B1 eastern Beata Ridge) (Figs. 1c, 11). In these record sections, it is possible to observe that the A1 section does not show a loss of energy while B1 section does show such loss. This fact could also be due to the interaction between Beata Ridge and Muertos Trough, and its effect into the eastern area of Hispaniola Island.

Crossing this structure, they remain constant at 8 km depth due to the fact that this seismic profile crosses the north-western Venezuela Basin, which at this is named as ''Dominican sub-basin''. This basin is characterized by two reflection horizons, A″ and B″ (Fig. 10). The next layer is located at 10 km of depth

Table 1

*Travel time fit for each phase*

| Phases | NC | Sigma (s) | NR | Trms (s) | $v^2$ |
|---|---|---|---|---|---|
| $P_2/P_2P/P_2'$ | 126 | 0.098 | 120 | 0.100 | 1.144 |
| $P_3/P_3P/P_3'$ | 250 | 0.122 | 224 | 0.110 | 0.896 |
| $P_4/P_4P/P_4'$ | 555 | 0.118 | 535 | 0.120 | 1.066 |
| $P_5/P_MP$ | 428 | 0.131 | 406 | 0.121 | 0.951 |
| $P_n$ | 139 | 0.173 | 139 | 0.128 | 0.466 |
| $P_6P$ | 184 | 0.163 | 184 | 0.142 | 0.850 |
| $P_7P$ | 57 | 0.143 | 57 | 0.133 | 0.437 |
| $P_9P$ | 46 | 0.150 | 46 | 0.126 | 0.955 |
| FIT | 1785 | 0.130 | 1711 | 0.121 | 0.951 |

*NC* number of picks, *Sigma* mean pick uncertainty, *NR* number of traced rays; *Trms* travel time root mean square misfit, $v^2$ normalized Chi square

Table 2

*Calculation of Moho boundary depth accuracy from the comparison between modeled and varied lower crust velocity (LCV) in km/s*

|  | Haiti basin | Beata island | Beata ridge | Total average (km) |
|---|---|---|---|---|
| LCV (km/s) | 0–50 (km) | 50–110 (km) | 120–170 (km) |  |
| 6.7 | −0.31 | −1.02 | −0.93 | −0.73 |
| 7.0 | 0.15 | 0.25 | 0.21 | 0.18 |
| 7.2 | 1.57 | 3.23 | 0.75 | 1.43 |

The values represent the differences between Moho depth of P-wave velocity model proposed in this paper and Moho depths obtained from different LCVs along three model regions: Haiti Basin, Beata Island and Beata Ridge with a P-wave velocity of 5.7–6.5 ± 0.1 km/s. The thickness of this layer beneath Beata Island is 10 km. The depth uncertainty calculations for shallow part reveal values less than 1 km. The shallowest part of the crust and seafloor in Beata Ridge region are characterized by recent sediments, which have been deformed by tectonic processes, as well as basement highs and mounds on the seafloor (e.g., MASSON and SCANLON 1991; DRISCOLL and DIEBOLD 1999; or MAUFFRET et al. 2001; among others). Along this model (Fig. 10), it is possible to observe that lower crust is more homogeneous than shallow part. The P-wave velocities vary from 6.6 to 6.9 km/s in the areas of Haiti sub-basin and Beata Ridge, respectively.

The Moho discontinuity is found to be deeper under Beata Island with a maximum value of 24 ± 2 km, thus confirming that the crust of Caribbean Plateau is unusually thick, being thinned continental or transitional type. To the east of Beata Island, the crust could be thickened oceanic crust type modified by the plateau, whereas the Moho depth rises up to 13 km in the area of seamount that is located further east, and then remains constant at 15 km in the Dominican sub-basin, typical of an oceanic crust. Deeper layers follow the Moho topography with a $V_P$ increasing in depth. In this study, it has been characterized one layer and two floating reflectors in the uppermost part of mantle, whose maximum velocities reach values of 8.5 ± 0.2 km/s. South to LM2 seismic profile, GRANJA-BRUÑA et al. (2014) suggest that Haiti sub-basin is 5 km thick. However, our study does not determine exactly the crustal thickness but we can affirm that is thicker than 5 km.

Previous wide-angle seismic studies of this area have provided information on the southern part of this zone. FOX et al. (1970) collated some previous studies in the Beata Ridge area. Some of them used refraction data and estimated the sedimentary cover to be 2 km thick, with P-wave velocities in the range 1.9–4.2 km/s, which are characterized by two prominent horizons. Moreover, the upper crustal layer over the Beata Ridge is 3–5 km thick, with a

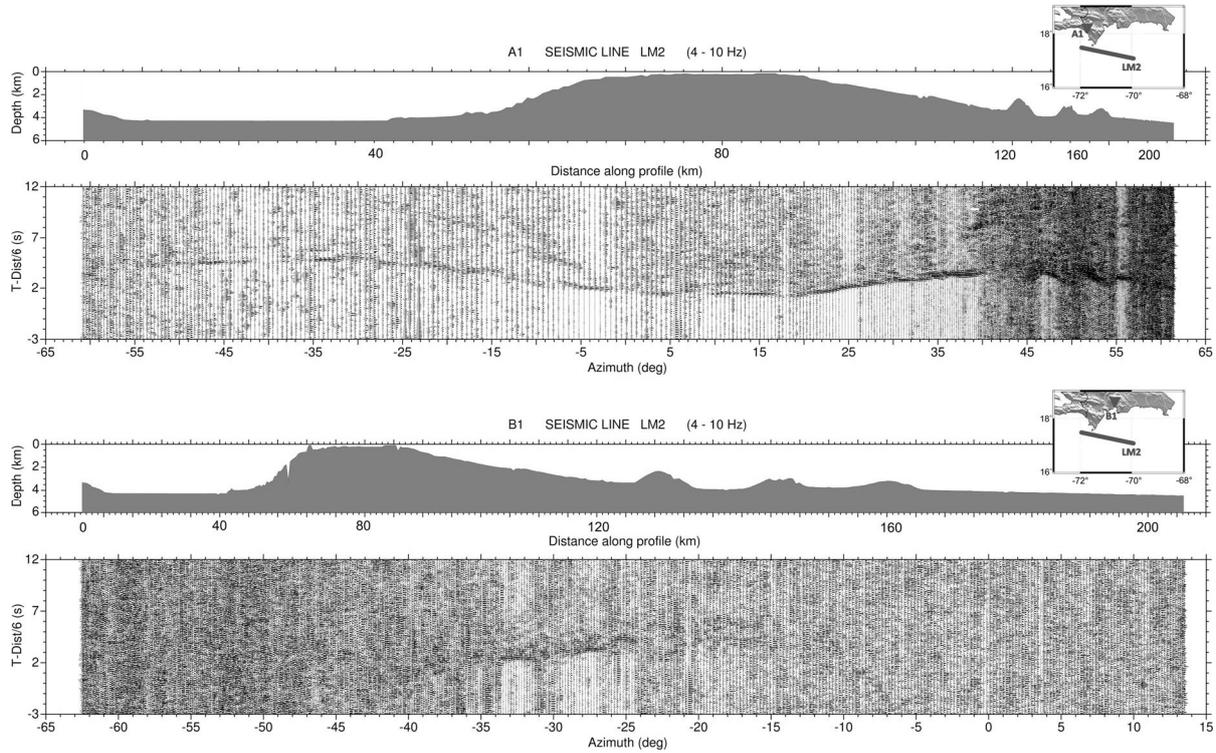

Figure 11

*Top* Azimuthal vertical seismic section of station A1 registering seismic line LM2; *Bottom* Azimuthal vertical seismic section of station B1 registering seismic line LM2. Bathymetry is shown on *top* of *both lines*. In the *upper right corner*, *triangle* shows location of registration point and *marine shooting line in thick line*. Reduction velocity is 6 km/s. Band-pass filtering of 4–10 Hz, was applied to both sections. Amplitudes are trace normalized compressional wave velocity of 5.4–5.9 km/s, whereas the lower crustal layer has a velocity of 6.7 km/s and a thickness of 7–10 km.

It is noteworthy that the P-wave velocities obtained in this region are in agreement with those proposed by DRISCOLL and DIEBOLD (1999). They established that upper crustal velocities tend to be approximately 5.5 km/s, although occasionally a thin upper layer is detected with velocities from 4.2 to 4.9 km/s. Mid-crustal velocities of 6.6 km/s are typical, and often velocities around 7.1–7.2 km are seen below these. By these authors, 'normal' Moho velocities are never observed, but instead, critically refracted energy with velocities between 7.4 and 7.8 km/s appear, which corresponds to rays turning at depths below the Moho. According to JAMES (2007), the Caribbean has 4 km of sediments underlain by a thicker crust with velocity 6.1–6.5 km/s and a major discontinuity below the crust where the velocity is 7.4 km/s. These values are in agreement with crustal thicknesses and upper crustal velocities proposed for many oceanic plateaus, and, in concrete, for the Caribbean plateau (JAMES 2009).

The P-wave velocities and thicknesses obtained in this paper are close to those mentioned below, even though this study crosses an area to the north of those zones that were previously studied. Moreover, this study provides seismic discontinuities in the upper mantle never obtained in previous studies.

The thickness of the crust beneath the northern Beata Ridge provided by CARIBE NORTE seismic profile LM2 is not as deep as expected in western and eastern flanks, although crustal thickness increases beneath Beata Island. The differences in the eastern part of the model could be due to the existence of a contemporary set of active faults.

The extension of the Beata Ridge structure towards the NE direction is identified as a first-order limit between mountain and plain morphostructures that are present on Hispaniola Island (COTILLA *et al.* 2007). This morphotectonic alignment is transversely articulated by the EPGFZ and SFZ active faults.

## 8. Conclusions

The wide-angle seismic data interpreted along the northern Beata Ridge have provided new information about the crustal structure of this region. The modeling of these data has allowed to obtain important differences between Haiti sub-basin, Beata Island, Beata Ridge and Dominican sub-basin.

- Sedimentary cover has been well determined from wide-angle seismic data. The sediments are thicker in the Haiti sub-basin area, disappearing on Beata Island and reappearing between the seamounts of Beata Ridge. In Venezuela Basin area (Dominican sub-basin), it is practically flat. These seismic layers can be correlated with Horizons A" and B", characteristic reflectors of Venezuela Basin, which are related to the Middle to Early Miocene chalks and the respective Santonian to Conician basalts intervals.
- Strong deformation appears in the upper crust of Beata Ridge region and it is practically imperceptible in the middle and lower crust.
- The Moho topography has been characterized for the first time in this area. It has a pronounced thickening under Beata Island, indicating that is a thinned continental or transitional crust. Whereas, to the east, the crust is oceanic type but below the seamounts is thickened while in the area of further east Beata Ridge is thinned. The crust of Dominican sub-basin is typically oceanic. In the area of Haiti sub-basin, the crustal thickness is more than 5 km. Previous MCS data had not been able to establish these depths.
- This study has allowed to characterize seismically one layer and two floating reflectors in the upper mantle. Their maximum depth has been determined to be between 50–55 ± 3 km with P-wave velocities increasing in depth.
- The analysis of CARIBE NORTE record sections reveals a decrease in the seismic energy of Beata Ridge region. The azimuthal record sections of the two seismic land stations, located on the both sides of Beata Island, corroborate that this loss of seismic energy only appears to the east of Beata Island. The explanation could be the interaction between Beata Ridge, Muertos Trough and eastern Hispaniola.
- Comparisons with previous refraction and MCS studies establish a correspondence with the shallow crustal structure results, but not with the deep structure in the northern flank of Beata Ridge.

*Acknowledgments*

This paper is a contribution to the projects CTM2006-13666-C02-02, complementary action (CTM2008-02955-E/MAR) and TSUJAL (CGL2011-29474-C02-01). We express special thanks to the Captain, officers, and crew of the R/V Hespérides, the technicians of the Unidad de Tecnología Marina and the CARIBE NORTE working group. We are also grateful for their collaboration with Dirección General de Minería, Universidad Autónoma de Santo Domingo (UASD), Instituto Sismológico Universitario (ISU) and Marina de Guerra Dominicana who provided two coastguard vessels 106-BELLATRIX and 109-ORION. D. Nuñez was funded with a doctoral grant from the Spanish Ministry of Education and Science. We thank to two anonymous reviewers for their helpful comments on the manuscript.

REFERENCES

AFILHADO, A., MATIAS, L., SHIOBARA, H., HIRN, A, MENDES-VICTOR, L., and SHIMAMURA, H. (2008), *From unthinned continent to ocean: The deep structure of the West Iberia passive margin*, Tectonophysics, *458*, 9–50.

BURKE, K., FOX, P., and SENGÖR, A.M.C. (1982), *Buoyant ocean floor and the evolution of the Caribbean*, J. Geophys. Res., *83*, 3949–3954.

BURKE, K. (1988), *Tectonic evolution of the Caribbean*. Annual Rev. Earth and Planetary Science Letter, *16*, pp. 201–230.

CARBÓ, A., CÓRDOBA, D., MARTÍN-DÁVILA, J., GRANJA-BRUÑA, J.L., LLANES, P., MUÑOZ-MARTÍN, A., and TEN BRINK, U. (2010),

Exploring active tectonics in the Dominican Republic, EOS, vol. *91*, no. 30, 261–268.

CATALÁN, M., and MARTÍN-DÁVILA, J. (2013), *Lithospheric magnetic mapping of the northern Caribbean region*. Geologica Acta, Vol. *11*, n. 3, pp. 311–320.

CERVENY, V., MOLOTKOV, I., and PSENCIK, I. (1977), Ray Method in Seismology. Prague, Czechoslovakia: University of Karlova.

COTILLA, M.O., CÓRDOBA, D., and CALZADILLA, M. (2007), *Morphotectonic Study of Hispaniola*, Geotectonics, vol. *41*, No. 5, pp. 368–391.

DIEBOLD, J. B., DRISCOLL, N. W. and EW-9501 SCIENCE TEAM, New Insights on the Formation of the Caribbean Basal Province Revealed by Multichannel Seismic Images of Volcanic Structures in the Venezuelan Basin, in: HSÜ, K.J., (Series Ed.). Sedimentary Basins of the World, 4. Caribbean Basins. MANN, P. (Ed), Elsevier Science, (N. Y., 1999), pp. 591–626.

DIEBOLD, J.B., Submarine volcanic stratigraphy and the Caribbean LIP's formational environment. In: JAMES, K.H., LORENTE, M.A., PINDELL, J.L. (Eds.), The Origin and Evolution of Caribbean Plate. Geological Society, (London, 2009), Special Publications, 328, pp. 799–808.

DIEHL, T., KISSLING, E., HUSEN, S., ALDERSONS, F. (2009), *Consistent phase picking for regional tomography models: application to the greater Alpine region*, Geophys. J. Int., vol. *176*, pp. 542–554.

DONNELLY, T.W., MELSON, W., KAY, R., and ROGERS, J.J.W. (1973), *Basalts and dolerites of late Cretaceous age from the Central Caribbean*. Init. Rep. DSDP. Washington, D.C., Government Printing Office *15*: 989–1012.

DONNELLY, T.W., BEETS, D., CARR, M.J., JACKSON, T., KLAVER, G.T., LEWIS, J., MAURY, T., et al., History and tectonic setting of Caribbean magmatism. In: DENGO, G., and CASE, J.E., eds. The geology of North America. (Boulder, Colorado, 1990). Geol. Soc. Am., pp. 339–374.

DRISCOLL, N. W., and DIEBOLD, J. B. (1998), *Deformation of the Caribbean region: One plate or two?*, Geology, *26*, 1043–1046.

DRISCOLL, N. W., and DIEBOLD, J. B., Tectonic and stratigraphic development of the eastern Caribbean: new constraints from multichannel seismic data, in: HSÜ, K.J., (Series Ed.). Sedimentary Basins of the World, 4. Caribbean Basins. MANN, P. (Ed), Elsevier Science, (N. Y., 1999), pp. 591–626.

EDGAR, T., EWING, J. and HENNION, J. (1970), *Seismic refraction and reflection in the Caribbean Sea*, Bull. Am Assoc. Petrol. Geologists.

EWING, J., ANTOINE, J., and EWING, M. (1960), *Geophysical investigations in the Eastern Caribbean: Trinidad Shelf, Tobago Trough, Barbados Ridge, Atlantic Ocean*, Geol. Soc. Am. Bull., *68*, 897–912.

FOX, P.J., RUDDIMAN, W.F., RYAN, W.B.F., and HEEZEN, B.C. (1970), *The Geology of the Caribbean Crust, I: Beata Ridge*, Tectonophysics, *10*, 495–513.

GHOSH, N., HALL, S., and CASEY, J. (1984), *Sea floor spreading magnetic anomalies in the Venezuelan Basin: the Caribbean-South American Plate Boundary*. Geol. Soc. Am. Mem. *162*:65–80.

GRANJA BRUÑA, J.L., TEN BRINK, U., CARBÓ-GOROSABEL A., MUÑOZ-MARTÍN, A., and GÓMEZ, M. (2009), *Morphotectonics of Central Muertos thrust belt and Muertos Trough (north-eastern Caribbean)*. Mar. Geol. *263*: 7–33.

GRANJA BRUÑA, J.L., CARBÓ-GOROSABEL, A., LLANES ESTRADA, P., MUÑOZ-MARTÍN, A., TEN BRINK, U.S., GÓMEZ BALLESTEROS, M., DRUET, M., and PAZOS, A. (2014), *Morphostructure at the junction between the Beata ridge and the Greater Antilles island arc (offshore Hispaniola southern slope)*, Tectonophysics, *618*, 138–163.

International Seismological Centre, On-line Bulletin, http://www.isc.ac.uk, Internatl. Seis. Cent., Thatcham, United Kingdom, 2001.

JAMES, K.H., (2007), The Caribbean Ocean Plateau—an overview, and a different understanding (http://kjgeology.com/wp-content/uploads/2014/01/2007-the-CaribbeanPlateau.pdf).

JAMES, K.H., In situ origin of the Caribbean: discussion of data, In: JAMES, K.H., LORENTE, M.A., PINDELL, J.L. (Eds.), The Origin and Evolution of Caribbean Plate. Geological Society, (London, 2009), Special Publications, *328*, pp. 77–125.

MANN, P., F. W. TAYLOR, R. L. EDWARDS, and KU, T.L. (1995), *Actively evolving microplate formation by oblique collision and sideways motion along strike-slip faults: An example from the northwestern Caribbean plate margin*, Tectonophysics, *246*, 1–69.

MANN, P., E. CALAIS, J.C. RUEGG, C. DEMETS, P. E. JANSMA, and G. S. MATTIOLI, (2002), *Oblique collision in the northeastern Caribbean from GPS measurements and geological observations*, Tectonics, *21*(6), 1057, doi: 10.1029/2001TC001304.

MASSON, D.G., SCANLON, K.M., (1991). *The neotectonic setting of Puerto Rico*. Geol. Soc. Am. Bull. *103*, 144–154.

MAUFFRET, A., and LEROY, S. (1997), *Seismic stratigraphy and structure of the Caribbean igneous province*, Tectonophysics, *283*, 61–104.

MAUFFRET, A., and LEROY, S., Neogene intraplate deformation of the Caribbean plate at the Beata Ridge, In: *Sedimentary Basins of the World*, vol. 4, *Caribbean Basins*, edited by P. Mann, (Elsevier Sci., 1999) pp. 627–669.

MAUFFRET, A., LEROY, S., VILA, J.-M., HALLOT, E., MERCIER DE LEPINAY, B., DUNCAN, R.A. (2001), *Prolonged magmatic and tectonic development of the Caribbean Igneous Province revealed by a diving submersible survey*. Mar. Geophys. Res. *22*, 17–45.

MERCIER DE LEPINAY, B., MAUFFRET, A., JANY, I., BOUYSEE, P., MASCLE, A., RENARD, V., STEPHAN, J.F. and Hernandez, E., (1988), *Une collision oblique sur de la bodure nord-caraïbe à la junction entre la ride de Beata et la fosse de Muertos*. C.R. Acad. Sci, Paris, *307*:1289–1296 (in French).

MOORE, G.T., and FAHLQUIST, D.A. (1976), *Seismic profile tying Caribbean DSDP Sites 153, 151, and 152*, Geol. Soc. Am. Bull., v. *87*, p. 1609–1614.

NÚÑEZ, D., KISSLING, E., CÓRDOBA, D., and PAZOS, A., (2011), *Consistency in CSS phase correlation: Application to CARIBE NORTE data set*, Geophysical Research Abstracts, vol. *13*, EGU2011–632, EGU General Assembly 2011.

RÉVILLON, S., HALLOT, E., ARNDT, N.T., CHAUVEL, C., and DUNCAN R. A. (2000), *A Complex History for the Caribbean Plateau: Petrology, Geochemistry, and Geochronology of the Beata Ridge, South Hispaniola*. J. Geol., vol. *108*, p. 641–661.

TALWANI, M., WINDISCH, C.C., STOFFA, P.L., BUHL, P. and HOUTZ, R.E., Multichannel seismic study in the Venezuelan basin and Curacao Ridge. In: M. TALWANI and W.C.I. PITMAN (Eds.), Islands Arcs, Deep Sea Trenches, and Back-Arc Basins. Am. Geophys. Union, (Maurice Ewing Ser., 1977), *1*, pp. 83–98.

ZELT, C.A., and ELLIS, R.M. (1988), *Practical and efficient ray tracing in two-dimensional media for rapid traveltime and amplitude forward modeling*. Canadian Journal of Exploration Geophysics, Vol. *24*, NO. 1, pp. 16–31.

ZELT, C.A., and SMITH, R.B. (1992), *Seismic traveltime inversion for 2-D crustal velocity structure, Geophys. J. Int., 108*, 16–34.

ZELT, C.A., and FORSYTH, D.A., (1994), *Modelling wide-angle seismic data for crustal structure—Southeastern Grenville Province.*, J. Geophys. Res., Solid Earth, *99* (B6), 11687–11704.

ZELT, C.A. (1999), *Modelling wide-angle traveltime data. Geophys. J. Int.* *139*, 183–204.